\documentclass{iopart}

\usepackage{graphicx}
\usepackage{psfrag}

\graphicspath{{eps/},{~/investigacion/tesis/eps}}

\renewcommand{\>}{\rangle} 
\newcommand{\<}{\langle}
\newcommand{\mcU}{\mathcal{U}}
\newcommand{\mcO}{\mathcal{O}}
\newcommand{\mcI}{\mathcal{I}}
\newcommand{\mcH}{\mathcal{H}}

\newcommand{\rmH}{\mathrm{H}}
\newcommand{\rmq}{\mathrm{q}}

\newcommand{\unam}{Universidad Nacional Aut\'onoma de M\'exico, M\'exico}

\newcommand{\cic}{Centro Internacional de Ciencias, Cuernavaca, M\'exico}
\newcommand{\icf}{Instituto de Ciencias F\'{\i}sicas, \unam}
\newcommand{\newo}{Department of Physics, Tulane University, New Orleans, USA}
\begin{document}
\title{A trivial observation on time reversal in random matrix theory}
\author{L. Kaplan$^1$, F. Leyvraz$^{2}$, C. Pineda$^{2,3}$, and T. H. Seligman$^{2,3}$}
\address{$^1$\newo}
\address{$^2$\icf}
\address{$^3$\cic}
\ead{carlospgmat03@gmail.com, seligman44@yahoo.com.mx}

\begin{abstract}
It is commonly thought that a state-dependent quantity, after
being averaged over a classical ensemble of random Hamiltonians,
will always become independent of the state. We point out that this
is in general incorrect: if the ensemble of Hamiltonians
is time reversal invariant, and the quantity involves the state in 
higher than bilinear order, then we show that the quantity is only 
a constant over the orbits of the invariance group on the Hilbert 
space. Examples include fidelity and decoherence in appropriate 
models. 
\end{abstract}


Whereas eigenfunctions of time reversal invariant (TRI) systems can always be
chosen real, complex linear combinations of such functions display different
statistical properties than real ones.  While this observation is trivial,
its consequences for applications of random matrix theory (RMT) have been
largely ignored. We shall show that these consequences are important in
properties of higher (mainly fourth) order in the wave functions, such as
transition probabilities, (inverse) participation ratios, fidelity, purity,
or von Neumann entropy. While these effects are often subleading in the
dimension $N$ of Hilbert space, for several quantities of physical interest,
including inverse participation ratios and purity decay rates, the
effects do appear at leading order.  Moreover, when considering entanglement
or decoherence in the context of quantum information we often deal with small
Hilbert spaces, possibly even a single qubit, where the choice of a real or
complex initial state becomes extremely important. Among the classical
ensembles of Hamiltonians, namely the Gaussian orthogonal, unitary, and
symplectic ensembles (GOE, GUE, GSE) as described by Cartan~\cite{cartanRMT},
GOE and GSE represent TRI systems and GUE represents non-TRI ones. In the GUE
case, averaging over Hilbert space is implicit in the ensemble average, but
this is not the case for the two TRI ensembles with important consequences,
some of which will be discussed in this paper. The same holds for the
corresponding circular ensembles of unitary matrices~\cite{mehta}. Among more
general ensembles, such as the chiral ones, similar distinctions have to be
made~\cite{guhr98random}. Note that the problems addressed involve statistics
of the wave functions only, and do not concern spectral properties. Thus, the
effects apply equally to TRI Gaussian and circular
ensembles. To avoid needless repetitions we limit ourselves to the GOE
and COE for the TRI case.

A very simple problem illustrates the type of effect we deal with.
Let us look at the autocorrelation function
\begin{equation}
A(t) = \vert \langle \psi | e^{- \rmi H t} |\psi \rangle \vert ^2
\end{equation}
for a TRI Hamiltonian $H$ drawn from a GOE. Its long time average is equal to
the inverse participation ratio $\mcI_{|\psi\>}=\sum_{|\alpha\>} |\<\alpha |
\psi\>|^4$, where $|\alpha\>$ are the eigenstates of $H$. Since
$\mcI_{|\psi\>}$ is a time average, we may hope that it is equal to an
ensemble average, independent of the initial state $|\psi\>$, for a chaotic
or mixing dynamics.  Yet one easily finds that $N \mcI=2N/(N+1) \stackrel{N
\to \infty}{\overrightarrow{\qquad}} 2$ when averaged over all (complex)
states, while $N \mcI=3 N /(N+2) \stackrel{N \to
\infty}{\overrightarrow{\qquad}} 3$ when averaged over all real states. The
result is obvious, since in the former case $\<\alpha | \psi\>$ behave as
random complex variables, while in the latter case they behave as random real
variables. Such effects on the inverse participation ratio have been studied
before in the context of wave function statistics in billiards with
TRI~\cite{biesstadium}.  There a comparison was made between the choice of
real or complex initial state in TRI systems and the choice of a symmetric or
non-symmetric initial state in a system possessing a unitary symmetry, such
as parity. Furthermore, initial states were discussed that are linear
combinations of real and complex random wave functions, exhibiting a
transition between the two limits.

In fact, taking a long-time average is unnecessary, and the effect is already
visible in the short-time dynamics of a TRI system. The average
autocorrelation function, for large $N$, is given by
\begin{equation}
\< A(t>0) \> =
\cases{
\frac{2-b_2(t/\tau_\rmH)}{N} & for complex $|\psi\>$,\\
\frac{3-b_2(t/\tau_\rmH)}{N} & for real $|\psi\>$}.
\end{equation}
$\< \cdots \>$  indicates an ensemble average,
$\tau_\rmH$ is the 
Heisenberg time of the Hamiltonian, and $b_2(t)$ is 
the two level form factor of the GOE~\cite{guhr98random}. 
For $0<t\ll \tau_\rmH$, we
have simply $\<A(t)\>\approx1/N$ and $\<A(t)\>\approx2/N$ 
respectively, i.e. the short-time
return probability for a real initial state is double that of a complex
initial state in a TRI system. This weak localization effect is easily
understood in semiclassical terms, since the factor of 2 results from
constructive interference between each returning path and its time-reversed
counterpart.

From these trivial examples we see immediately that whenever we average a
quantity that is not bilinear in the wave function, the average over a TRI
ensemble such as GOE or GSE (and thus the time average if ergodicity holds)
does depend on whether the initial state is real (up to an overall phase) or
complex.

Proposing an experiment is not altogether trivial. One possibility is to
excite a solid metal block elastically with $M$ pings at different times and
places, corresponding to a state $|\psi\>=\sum_{j=1}^M e^{i \phi_j}
|\psi_j\>$, where each $|\psi_j\>$ is real but the relative phases are
random. Assuming all pings have the same strength, the average
autocorrelation function at long times yields $(2+1/M)/N$, e.g. $2.5/N$ for
two pings versus $3/N$ for a single ping. Similarly, at short times we have
$(1+1/M)/N$. Such an experiment can be performed~\cite{weaverprivate2007},
though it might not be all too interesting as the outcome is clear.


On a slightly more formal note we may say the following: Starting from an
arbitrary state $|\psi\>$ in some Hilbert space $\mcH$ of dimension $N$, we
shall cover, up to normalization, the entire Hilbert space by the orbit of
$|\psi\>$ traced by  $\mcU (N)$ on  $\mcH$.

Consider now that the GUE is can be defined as a set of diagonal matrices
$\Delta$ with the appropriate measure $\rmd\nu (\Delta )$ composed with the unitary
matrices $u$ as $u^\dagger \Delta u$, with the invariant Haar measure
$\rmd\mu (u)$.  It is then immediately clear that averaging over the
GUE will include averaging over all states.  In other words, a state
dependent quantity averaged over the Hamiltonians is automatically constant
throughout the Hilbert space.  If, on the other hand, we consider a GOE, the
corresponding representation is $o^\dagger \Delta o$ with the Haar measure of
the orthogonal group $\rmd\mu (o)$.  The implicit averaging over states will
then be limited to the orbits of $\mcO(N)$ on the Hilbert space. A state
dependent quantity averaged over the Hamiltonians, in the TRI case, is
therefore only constant on the orbits of the original state. 

\begin{figure}
  \centering 
  \includegraphics{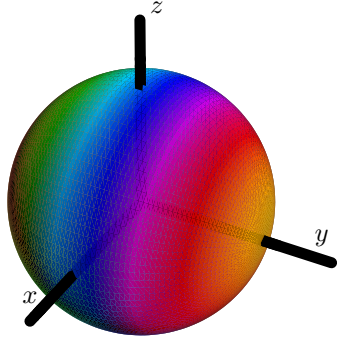} 
  \caption{The Bloch sphere~\cite{NC00a}, representing the orbits of states
    under the action of $\mcO(2)$.} 
  \label{fig:blochgoe}
\end{figure}

If we consider the circular ensembles, the situation is slightly more
involved, as the symmetry operations defining these ensembles are not
similarity transformations. Recall that the CUE is the unitary group
$\mcU(N)$ itself and thus is left {\it and} right invariant under $\mcU(N)$;
obviously this includes similarity transformations and thus again the
ensemble average includes state averaging. For the COE the situation is more
complicated as the measure is invariant under $\mcU(N)$, but if $S$ is an
element of the COE the operation is defined as $\rmd\mu (S) =\rmd\mu (u^t S
u)$.  Note that this is not a similarity transformation. Yet if we restrict
$\mcU (N)$ to $\mcO (N)$ we have a similarity transformation as the transpose
of an orthogonal matrix is its inverse.  Thus the same orbits discussed above
describe the averaging we achieve with the ensemble average of the Hamiltonians.
The two-dimensional Hilbert space associated with a qubit provides the best
example. Representing this space in terms of the Bloch sphere, the orbits of
$\mcO(2)$ are rings around the $y$ axis of the sphere as illustrated in
\fref{fig:blochgoe}.

Considering the recent interest in developing RMT models for fidelity
decay~\cite{0305-4470-36-12-334, 1367-2630-6-1-020, ShepelyanskyRMTFidelity}
and decoherence \cite{1464-4266-4-4-325, Gorin2003, reflosch, pinedaRMTshort,
pinedalong}, we may ask how the effect appears in this context. Clearly, the
effect will not be observed for the fidelity amplitude or for coherences
(off-diagonal elements of the density matrix), since these quantities are
bilinear in the wave functions. On the other hand, considerations regarding the
choice of initial state will be highly relevant in the study of fidelity,
purity, or von Neumann entropy. Indeed initially puzzling results on purity
decay for one or two qubits~\cite{pinedalong} lead to, and will be at the
centre, of the present analysis.  The importance of entanglement and
decoherence of one- and two-qubit systems as the basic building blocks of
quantum information~\cite{NC00a} justifies this focus, particularly as
manipulations of qubits using tools of quantum optics allow complex states to
be produced in a very simple way.  Yet we shall start with the simpler case of
fidelity decay, a benchmark in quantum information.

The fidelity amplitude is defined as
\begin{equation}
 f_\epsilon (t)= \<\psi |\rme^{-\rmi H_0 t}  \rme^{\rmi H_\epsilon t}|\psi\>
 \end{equation}
and fidelity is given as $F_\epsilon(t) =|f_\epsilon (t)|^2$, where
$H_\epsilon = H_0 + \epsilon V$. Here $|\psi\>$ is any function,  $H_0$ is
any Hamiltonian of interest, $V$ a perturbation, and $\epsilon$ a real
parameter.  In \cite{1367-2630-6-1-020} both $H_0$ and $V$ were modeled by
random matrix ensembles. While in some special cases exact results were
obtained \cite{1367-2630-6-1-199, shortRMT, gorin:244105}, linear response
results were obtained in a wide variety of cases (see \cite{reflosch} and
references therein).  In \cite{1367-2630-6-1-020, reflosch} the linear
response result for $F_\epsilon (t)$ is given as
\begin{equation}
	\<F(t)\> = \<|f_\epsilon (t)|^2\>  
           =  \<f_\epsilon (t)\>^2 +(2\pi \epsilon)^2 (2/\beta_V) \mcI \  t^2 +\Or(\epsilon ^4),
\end{equation}
where $\mcI$ is the inverse
participation ratio of $|\psi\>$ in the eigenbasis of $H_0$ and $\beta_V =
1,2,4$ specifies whether the perturbation is taken from a GOE, GUE, or GSE.
The validity of this approximation can be extended by exponentiating the
second order term.  The conclusion in these papers was that for a random
state $\mcI=\Or(1/N)$, and thus $F_\epsilon \approx \<f_\epsilon\>^2$. If we
further average over random states, this is still correct. Yet we can
reasonably ask what the state-averaged fidelity would be if the average is
limited to some small subspace of the total Hilbert space, which for
some reason is interesting or experimentally accessible. Then the
inverse participation ratio is large and we observe an
effect of order one exactly like the one discussed above.
The result without state average will depend on the initial state if
we allow complex states and consequently self averaging is lost.  In this
example the choice of the subspace over which we average might be somewhat
arbitrary, though often only certain frequency regions are accessible. 

For  a composite system, the situation is different. We often have a natural
separation between a smaller system, which we call the central system, and an
environment that interacts with it.  Typically we will only be interested in
the central system, or only the central system may be accessible to
experiment.  Such is the case for one or a few qubits coupled to an
environment, where both the environment and the coupling are modeled by
random matrix ensembles~\cite{pinedaRMTshort, pinedalong}. 

\begin{figure} \centering
\psfrag{P}[tl]{$P$} \psfrag{S}[b]{$S$} \psfrag{toth}{$t/\tau_\rmH$}
\psfrag{q}{$\log_2 N_\rme$}\psfrag{log}[tc]{$\log_2 \sigma_P$}
\includegraphics{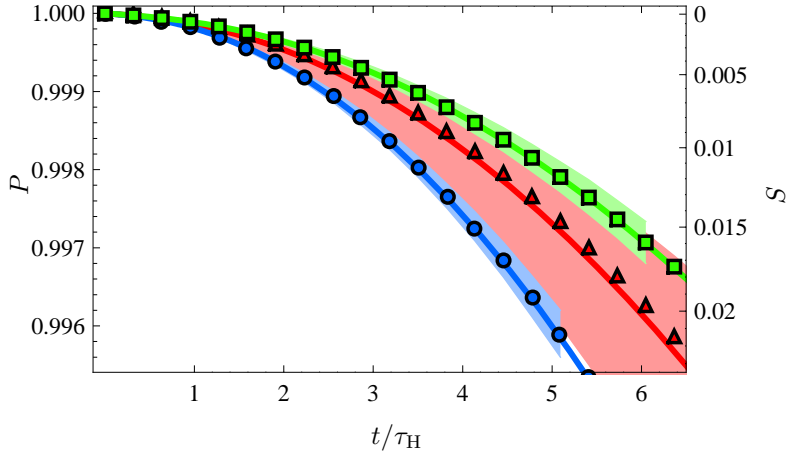}
\caption{We display the behaviour of purity and von Neumann entropy, for
  initial qubit states in different regions in the Bloch sphere, fixing
  $\lambda=10^{-3}$ and environment size $N_e=1024$. The coding is as follows:
  green (squares) for $\gamma=0$, blue (circles) for $\gamma=\pi/2$, and red
  (triangles) for arbitrary $\gamma$.  For each case, the calculation was
  repeated 100 times with randomly chosen realizations of $H_\rme$ and
  $V_{\rme,\rmq}$, and random initial states.  The shaded coloured regions
  represent envelopes encompassing all 100 runs, the average purity is
  indicated by symbols, and the predicted behaviour of \eref{eq:purityGOEsep}
  is shown by solid curves.}
  \label{fig:decayGOE}
\end{figure}

Purity of a density matrix $\rho$ is defined as $P(\rho)=\tr \rho^2$, and is
a measure of the degree of mixedness of the density matrix, or the degree of
entanglement of a central system with an environment. Thus it is also a
measure of decoherence.  Consider a single qubit and an environment evolving
under the Hamiltonian
\begin{equation}
	H=H_\rme + \lambda V_{\rme,\rmq}
\label{eq:onequbit}
\end{equation}
where $H_\rme$ acts on an environment of dimension $N_\rme$, $V_{\rme,\rmq}$
is a $2N_\rme\times 2N_\rme$ matrix coupling the qubit and the environment, and
$\lambda$ a parameter controlling the strength of the coupling. The initial
state is a pure separable state. Let us choose both $H_\rme$ and
$V_{\rme,\rmq}$ from the GOE. The resulting ensemble of Hamiltonians is
invariant under local orthogonal transformations.  Evaluating purity 
of the qubit density matrix, in
linear response approximation, we obtain for large $N_\rme$~\cite{pinedalong}:
\begin{equation}\label{eq:purityGOEsep} 
  P(t)=1-\lambda^2 \left\{ 
     t^2 \left[3-\cos(2\gamma)\right] +2 t \tau_\rmH - 2 B_2(t)  
             \right\}, 
\end{equation} 
where 
\begin{equation}\label{eq:Btwogoe} 
  B_2(t)= 
    2\int_0^t \rmd \tau \int_0^\tau \rmd\tau' b_2(\tau'/\tau_H) 
\end{equation} 
is the double integral of the two-level GOE form factor and
$\tau_\rmH$ is the Heisenberg time of the environment. $\gamma$ is the angle
between the $xz$ plane and the representation in the Bloch sphere of the
initial state of the qubit.  As, for a single qubit, purity and von
Neumann entropy $S=-\tr \rho \log \rho $ have a one to one relation,
\eref{eq:purityGOEsep} can be translated easily to obtain entropy increase. 

\begin{figure} \centering
\psfrag{P}[tl]{$P$} \psfrag{S}[b]{$S$} \psfrag{toth}{$t/\tau_\rmH$}
\psfrag{q}{$\log_2 N_\rme$}\psfrag{log}[tc]{$\log_2 \sigma_P$}
\includegraphics{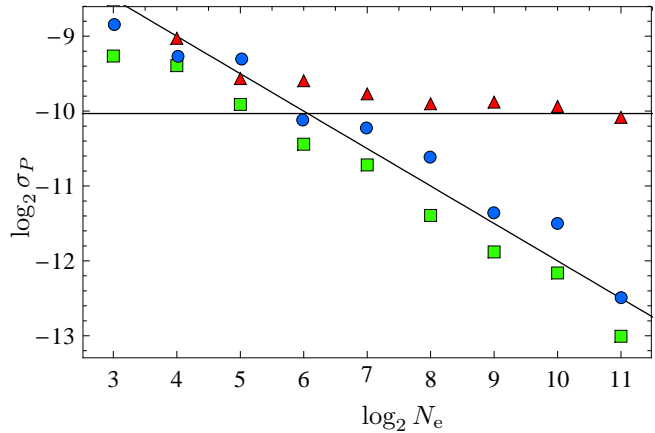}
\caption{We plot the standard deviation of purity $\sigma_P$ at time $t=40$,
  as a function of the environment dimension, using the same coding and value
  of $\lambda$ as in the previous figure. For a fixed value of $\gamma$ (blue
  circles 
  and green squares), there is asymptotic self averaging, as indicated by the line
  $\propto 1/\sqrt{N_\rme}$. In contrast, for arbitrary initial conditions
  (red triangles), the standard deviation at large $N_\rme$ approaches the finite value
  predicted in \eref{eq:sigmagoe}, here plotted as a horizontal line.}
  \label{fig:decaysigma}
\end{figure}

In \fref{fig:decayGOE} we show $P(t)$ for $\gamma= 0$ (green squares), for
$\gamma= \pi/2$ (blue circles), and for random initial states in the whole
Bloch sphere (red triangles).  In contrast to the GUE case, the average
purity depends on the initial state (via the angle $\gamma$).  The fastest
decay of purity is observed for $\gamma= \pi/2$, where the state is
orthogonal to its time reversal image.  The slowest decay is observed for
$\gamma= 0$, which characterizes TRI states.  In \fref{fig:decaysigma} we
show numerical results for the standard deviation of the purity as a function
of $N_\rme$, the dimension of the Hilbert space of the environment. We
consider the same cases as in \fref{fig:decayGOE}.  Note that $H_\rme$,
$V_{\rme,\rmq}$, and the initial 
state of the environment are randomly chosen from
their respective ensembles. We see that for fixed $\gamma$, the standard
deviation falls off as $N_\rme ^{-1/2}$.  By contrast, the standard
deviation converges to a finite value when $\gamma$ is unrestricted.
Since, for $N_\rme \to\infty$, the variations
in $\cos 2\gamma$ are the only source of purity fluctuations, the
standard deviation of the purity is 
\begin{equation}\label{eq:sigmagoe}
  \sigma_P=
    \frac{4}{3\sqrt{5}} \lambda^2 t^2 + \Or (\lambda^4,N_\rme^{-1}).  
\end{equation}
This value is plotted in \fref{fig:decaysigma}.

This result for a single qubit is especially significant in view of the fact
that the decoherence of several qubits can often (in the high purity
approximation) be reduced to the case of a single qubit~\cite{GPS-letter}.

Summarizing, we have shown that, for the general (non-TRI) case, averaging
over the ensemble of Hermitian Hamiltonians (GUE) implies a full average
over all states. For TRI systems, on the other hand, if a state-dependent
quantity is averaged over the ensemble of real Hamiltonians (GOE), it
will in general still depend on the orbit of the initial state under the
orthogonal group. We have shown that this actually happens, if the averaged
quantity depends on the state in higher than bilinear order. The variance of
fidelity, in particular, shows such behaviour, but it is of order $1/N$,
where $N$ is the dimension of the system, and thus often insignificant.
We have further displayed a specific TRI random matrix model for the 
decoherence of a qubit, for which the effect is of order one.

This work was supported in part by the U.S. National Science Foundation under
Grant No. PHY-0545390. We acknowledge support from the grants UNAM-PAPIIT,
DGAPA IN112507 and CONACyT 57334. CP was supported by DGEP. THS thanks H. A.
Weidenmuller for useful discussion.

\section*{References}
\bibliographystyle{unsrt}
\bibliography{paperdef,miblibliografia,specialbib}
\end{document}